\documentclass[acmsmall]{acmart}

\AtBeginDocument{%
  }

\setcopyright{acmlicensed}
\copyrightyear{2025}
\acmYear{2025}
\acmDOI{XXXXXXX.XXXXXXX}

\begin{document}

\title{Influence Operations in Social Networks}

\author{Javier Pastor-Galindo}
\affiliation{%
  \institution{Computer Systems Engineering Department, Universidad Politecnica de Madrid}
  \city{Madrid}
  \country{Spain}
}

\author{Pantaleone Nespoli}
\author{José A. Ruipérez-Valiente}
\affiliation{%
  \institution{Department of Information and Communications Engineering, University of Murcia}
  \city{Murcia}
  \country{Spain}
}

\author{David Camacho}
\affiliation{%
  \institution{Computer Systems Engineering Department, Universidad Politecnica de Madrid}
  \city{Madrid}
  \country{Spain}
}


\begin{abstract}
An important part of online activities are intended to control the public opinion and behavior, being considered currently a global threat. This article identifies and conceptualizes seven online strategies employed in social media influence operations. These procedures are quantified through the analysis of 80 incidents of foreign information manipulation and interference (FIMI), estimating their real-world usage and combination. Finally, we suggest future directions for research on influence operations.
\end{abstract}

\ccsdesc[500]{Security and privacy~Social network security and privacy}

\keywords{social networks, disinformation, misinformation}


\received{20 February 2007}
\received[revised]{12 March 2009}
\received[accepted]{5 June 2009}

\maketitle

\section{THE INFLUENCE CYBER BATTLEGROUND}
The rise of social media and other online communication channels has provided unprecedented reach and efficiency for intentional influence campaigns, having an important escalation from Russia's interference in the 2016 U.S. election. Governments are worried about these risks and social platforms seriously fight against the so-called coordinated inauthentic behavior (CIB) ~\cite{PASTORGALINDO2022161}.

In 2023, the landscape of influence operations saw a significant rise in detected Foreign Information Manipulation and Interference (FIMI) incidents, with the European External Action Service (EEAS) report~\cite{EuropeanUnionExternalAction2024} identifying 750 cases between December 1, 2022, and November 30, 2023, nearly doubling the previous year's figures. FIMI attacks were globally distributed, targeting 53 countries, with Ukraine, the United States and several European nations being the most affected. The content dissemination involved the coordinated use of websites and social media to create illusions of authentic discourse and obscure origins, with cross-platform reach as a standard practice. Unfortunately, this global threat is expected to be massively automated with generative artificial intelligence, as stated by OpenAI~\cite{Goldstein2023}, Microsoft~\cite{Microsoft2024} or Google~\cite{Google2025}.

Consequently, the study of online influence operations has grown in the last few years~\cite{gabriel2023inductive}. Existing case studies have focused on characterizing disinformation methods, inferring strategies from observed behaviors, assessing political impacts, analyzing interactions or identifying users likely to spread deceptive content~\cite{PASTORGALINDO2022161}, among others. Particularly, there is a clear tendency towards proposing unsupervised and supervised machine learning methods for detection and analysis. However, the former are limited in scalability and timeliness, while the latter do not fully address variability and detectability over time and across campaigns~\cite{Alizadeh2020}. That is why the conclusions reported in the literature usually have a partial, biased, or limited understanding of the problem~\cite{Budak2024}. 

In this context, apart from existing applied research and empirical studies, more analytical research is needed to abstract the concept of an influence operation, set academic foundations and homogenize the vision of this global challenge~\cite{9451574}. Particularly, the interpretation of influence operations stands multidisciplinary, with several dimensions coming into play such as psychology, sociology, political science, communication studies or computer science. That is why there is little consensus on how to best describe and address influence operations~\cite{blazek2021scotch}. In this line, and recognizing that systematic modeling of influence operations is far from established, particularly on social networks, we propose seven broader influence strategies that allow for a unified and structured analysis of campaigns across social platforms.

\section{INFLUENCE OPERATIONS: TOWARDS A COMMON UNDERSTANDING}

Influence operations are deliberate activities performed by threat actors to sway public opinion and manipulate decision-making globally~\cite{Budak2024}. Their analysis is crucial for interpretation, standardized knowledge, and shareable intelligence, often guided by models adopted by agencies, organizations, and researchers~\cite{canovas2024analyzing}.

Several frameworks can aid in structuring influence operation analysis. The \textit{ALERT} taxonomy~\cite{DESOUZA2020101606} categorizes political disruption tactics, while the \textit{ABCDE} framework~\cite{pamment2020eu} assists EU institutions and platforms in identifying key operational components. The Carnegie Mellon \textit{BEND} framework~\cite{carley2020social} details structural and narrative maneuvers, and the U.S. government’s \textit{SCOTCH} framework~\cite{blazek2021scotch} integrates technological, organizational, and human factors for adversarial operations. One promising perspective is the consideration of them as cyber operations, calling for efforts to understand the problem from the cybersecurity discipline. The Norwegian Defense Research Establishment adapted the \textit{cyber kill chain} for social media influence campaigns~\cite{bergh2020understanding}, mapping the phases from reconnaissance to impact. Researchers at NYU~\cite{mirza2023tactics} developed a similar framework based on interviews with fact-checkers and analysts, characterizing threat actors, attack patterns, channels, and targets. More rigorous proposals have been proposed by the Misinfosec Working Group’s \textit{AMITT} framework~\cite{Walker2019}, later enhanced by MITRE and Florida International University into \textit{SP!CE}, extending MITRE ATT\&CK to influence campaigns. In 2022, these merged into the \textit{DISARM framework}~\cite{terp2022disarm}, now supported by the DISARM Foundation. While \textit{SP!CE} remains used by the U.S. Department of Defense, MITRE continues collaborating with \textit{DISARM} to counter cyber-enabled influence operations~\cite{Sixto2023}. Finally, Meta researchers introduced the \textit{Online Operations Kill Chain}~\cite{nimmo2023phase}, detailing ten steps used to analyze takedowns of influence campaigns on Facebook. 

While the reviewed frameworks effectively characterize general online operations, they lack a clear focus on understanding adversary campaigns on social media. Nevertheless, they provide a foundation to propose key strategies of influence operations in social networks. In particular, the \textit{DISARM} framework~\cite{terp2022disarm} is one of the most comprehensive and up-to-date alternatives, being currently employed in several projects (e.g., Attribution Data Analysis Countermeasures Interoperability~\cite{adacdisarm}, ATHENA~\cite{athenea} or the European Digital Media Observatory~\cite{edmoelections}). It has also been adopted by important organizations (such as EU DisinfoLab~\cite{disinfolabeufimi}, the European Center of Excellence for Countering Hybrid Threats (Hybrid CoE)~\cite{HybridCoE2022}, the European Union Agency for Cybersecurity (ENISA)~\cite{enisathreat}, the European External Action Service (EEAS)~\cite{EuropeanUnionExternalAction2024} or the Alliance for Securing Democracy~\cite{reportasd}), demonstrating its practical utility and reliability. The \textit{DISARM framework}\footnote{\url{https://disarmframework.herokuapp.com/}} follows the philosophy of the MITRE ATT\&CK matrix\footnote{\url{https://attack.mitre.org/}} and considers that influence operations are structured hierarchically into phases, tactics, and techniques. It defines the following four phases: 1) \textit{plan} (set clear goals and strategies), 2) \textit{prepare} (assemble resources for readiness, including people, networks, channels and content), 3) \textit{execute} (deploy the operation from start to finish, maintaining presence as needed), and 4) \textit{assess} (review the outcome to improve future maneuvers). As detailed in Table~\ref{tab:disarm}, each phase aggregates a set of tactics. Finally, each tactic can be achieved through specific fine-grained actions, called techniques. There are more than one hundred predefined techniques under the sixteen tactics, which can be explored in detail through the DISARM Framework Explorer\footnote{https://disarmframework.herokuapp.com/technique/}. For instance, in the phase of \textit{execute}, the tactic \textit{deliver content} includes techniques like \textit{deliver ads}, \textit{post content}, \textit{comment or reply on content}, and \textit{attract traditional media}. 

The resulting matrix is a robust tool for modeling and evaluating various types of influence operations in social networks.  Among other benefits, this cybersecurity-inspered modeling facilitates the understanding of adversary strategies, analyzing indicators of compromise (IOC), building tailored reactions to the precategorized techniques, increasing cooperative and collective resilience, and countering complex problems such as hybrid threats, advanced persistent threats (APT), or the last tailored term of advanced persistent manipulators (APM)~\cite{EuropeanUnionExternalAction2024,Microsoft2024}. 

\begin{table}[t!]
\footnotesize
\begin{tabular}{cccc}
\toprule
\textbf{Plan} & \textbf{Prepare} & \textbf{Execute} & \textbf{Assess} \\ \midrule
Plan strategy & Develop narratives & Conduct pump priming & Assess effectiveness \\ 
Plan objectives & Develop content & Deliver content &  \\
Target audience analysis & Establish social assets & Maximize exposure &  \\ 
& Establish legitimacy & Drive online harms &  \\
& Microtarget & Drive offline activity &  \\
& Select channels and affordances & Persist in the information environment &  \\ \bottomrule
\end{tabular}
\caption{DISARM phases and associated tactics in adversary influence operations.}
\end{table}
\label{tab:disarm}

\section{INFLUENCE OPERATIONS STRATEGIES IN SOCIAL NETWORKS}

\begin{figure*}[t!]
    \centering
    \includegraphics[width=\textwidth]{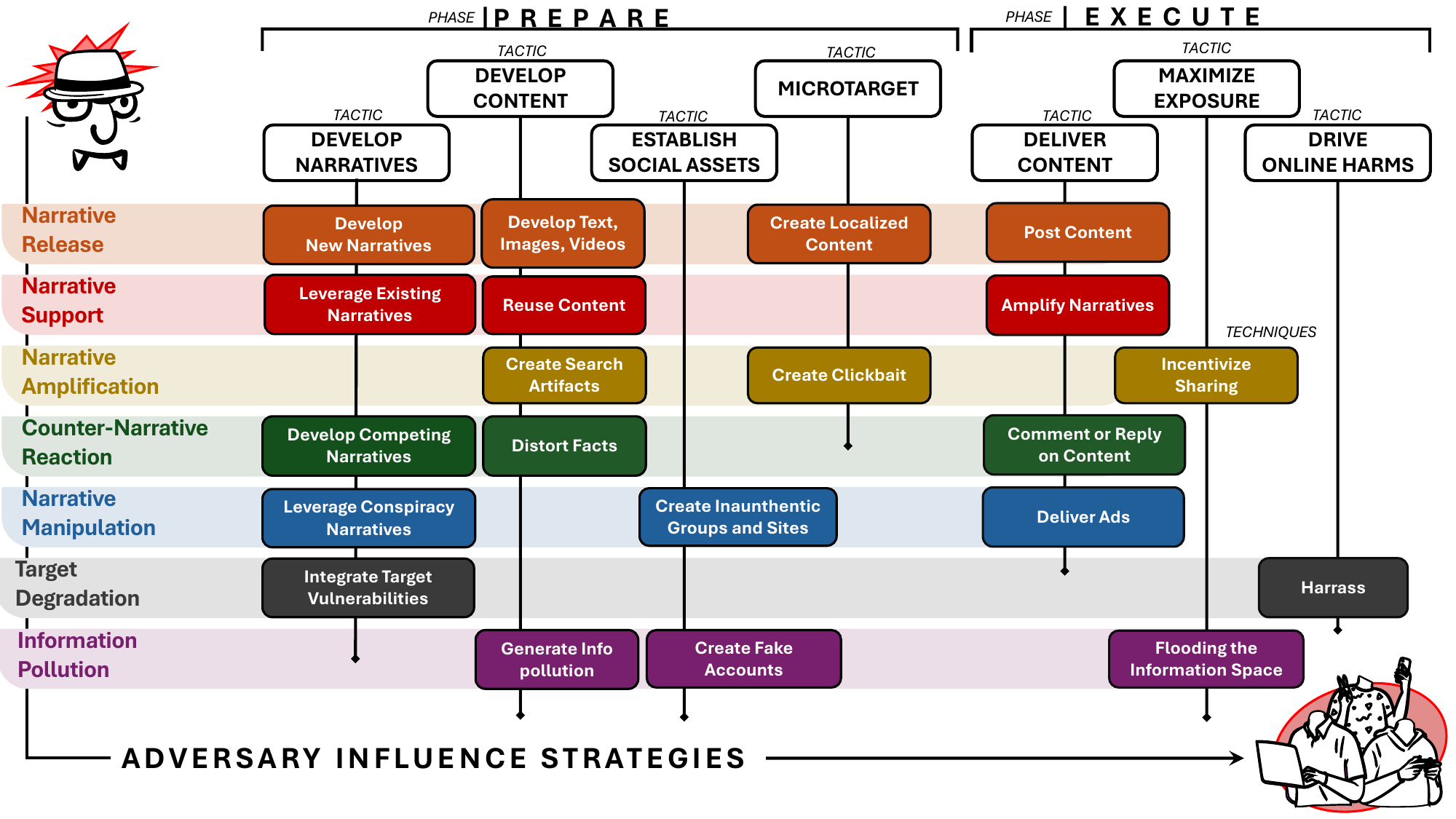}
    \caption{Seven strategies based on DISARM techniques for influence operations in social networks}
    \label{fig:disarm_lite}
\end{figure*}

Strategy refers to high-level procedures aimed at achieving long-term objectives. In influence operations on social networks, strategy represents a calculated approach to manipulate beliefs, perceptions, and behaviors. To address the lack of clear definitions in the literature, we infer influence strategies from the \textit{DISARM framework} as follows:

\begin{enumerate}
\item \textit{Identify execution-phase DISARM techniques specific to social networks}. While the \textit{DISARM} phases of plan, prepare, and assess are medium-agnostic, execution techniques are tied to communication channels. We select and focus on seven techniques exploiting social networks as the primary attack vector, shown in the right part of Figure~\ref{fig:disarm_lite}: \textsf{(1) Post content}, \textsf{(2) Amplify Narratives}, \textsf{(3) Incentivize Sharing}, \textsf{(4) Comment or reply on content}, \textsf{(5) Deliver ads}, \textsf{(6) Harass}, and \textsf{(7) Flood the Information Space}.  

\item \textit{Map execution techniques to their predecessor techniques from previous phase}. This reverse mapping exposes the strategic decisions and resource allocation required for executing influence strategies. By traversing the DISARM matrix, we form disjoint groups of techniques to define decoupled strategies. Consequently, the planning and assessment phases are excluded due to their universal applicability, while certain preparation and execution tactics are omitted as they do not serve as clear preconditions for the seven execution techniques.

\item \textit{Define influence strategies from resulting pipelines}. The sequential flow from preparation to execution forms structured activity pipelines, outlining influence patterns of threat actors. Each colored pipeline is inductively named and defined as a high-level influence strategy.  

\end{enumerate}

\subsubsection*{Narrative Release}
This strategy introduces a narrative into the social network that aligns with the intended influence objective. The fundamental technique is \textit{posting content}, which involves publishing original posts or multimedia directly on social platforms. These posts can be text updates, images, videos, or links. To enable the execution of this technique, adversaries must first develop narratives from scratch, develop the textual or multimedia content, and localize artifacts to ensure that messaging is culturally and linguistically adapted. For instance, an actor might post a publication on a Facebook group to shape public perception of a controversial issue. The goal is to establish an anchor point for influence, making the narrative recognizable and available for reinforcement, expansion, or manipulation~\cite{tyushka2022weaponizing}.

\subsubsection*{Narrative Support}
Once a narrative is introduced, this strategy strengthens its credibility and makes it more widely accepted. Adversaries rely on \textit{amplifying narratives} to make posts appear more popular than they naturally would be by leveraging existing narratives and reusing content. A common example is the forced posting or interaction with a narrative to increase reliability and engagement. The goal is to solidify public acceptance, making the narrative more resistant to scrutiny or opposing views~\cite{blumenstock2025migration}.

\subsubsection*{Narrative Amplification}
This strategy focuses on expanding the reach and virality of a narrative. By increasing exposure, the narrative appears more widespread and influential for a broader audience. Adversaries increase content spread by sharing and \textit{incentivizing sharing}, creating search artifacts that optimize content for visibility on search engines and social platforms while designing clickbait that exploits cognitive biases. For example, a well-crafted Instagram reel with provocative imagery and a catchy caption can encourage mass sharing. The goal is to raise its visibility, making it harder to ignore~\cite{hristakieva2022spread}.

\subsubsection*{Counter-Narrative Reaction}
Deployed in response to competing narratives, this strategy works to challenge, weaken, or redirect attention from opposing perspectives. Influence actors rely on \textit{commenting or replying on content} to shape discussions and control narratives. By strategically posting comments on public threads, they can reinforce messages, attack opposition, or create the illusion of widespread support. To execute this, adversaries develop competing narratives that counter opposition and distort facts to undermine credibility. This strategy is frequently used in YouTube comment sections, where adversaries flood replies with endorsements or criticism to sway public perception. The goal is to protect the dominant narrative, ensuring it remains the primary frame of reference~\cite{kaiser2022partisan}.

\subsubsection*{Narrative Manipulation}
Rather than simply reinforcing or amplifying a message, this strategy modifies, distorts, or falsifies existing narratives to reshape public perception. Adversaries use links, news articles, or social media promotions to redirect users to manipulated external content through \textit{delivering ads}. This strategy is supported by leveraging conspiracy narratives, creating inauthentic groups, and developing fake news sites as reference points for misleading claims. A Facebook ad campaign, for instance, might target users with deceptive links to fabricated news articles tailored to their biases. The goal is to redefine reality, aligning public perception with a preferred version of events~\cite{ren2022authoritarian}.

\subsubsection*{Target Degradation}
This strategy seeks to discredit, intimidate, or silence individuals or groups that challenge an influence operation. It diminishes their credibility with \textit{harassment}, weakening their influence, and discourages engagement to remove opposition from public debate. To prepare for such actions, adversaries integrate target vulnerabilities, gathering intelligence on their targets to personalize attacks. A common example is a Twitter harassment campaign targeting journalists or political figures to discredit or silence them. The goal is to reduce the ability of key figures to contest the dominant narrative~\cite{9451574}.

\subsubsection*{Information Pollution}
By overwhelming the information environment with excessive or conflicting content, this strategy creates confusion and distrust in the reliability of information sources. Adversaries can \textit{flood the information space}, not necessarily promoting a particular narrative but diluting clarity and credibility, flooding social networks with excessive content and making it difficult for users to discern credible information. This execution technique requires generating an excess of low-quality or misleading content and creating fake accounts to emit and amplify it. A typical example involves posting hundreds of tweets quickly to saturate a conversation and obscure opposing viewpoints. The goal is to increase uncertainty, leaving audiences more susceptible to influence~\cite{pan-etal-2023-risk}.

Overall, these strategies provide a generalized framework for understanding complex influence operations in the real world, being correlated with other studies~\cite{buchanan2021truth}.

\section{INFLUENCE OPERATIONS IN THE REAL-WORLD}
The literature on cross-analysis of influence operations using a standardized framework remains limited. However, over the past year, a dataset comprising 81 FIMI campaigns targeting elections and referendums from 2014 to 2024 has been manually tagged using DISARM techniques by the SIPA Institute of Global Politics (IGP)~\cite{FuldeHardy2024}, leveraging intelligence reports. By identifying the execution techniques employed in each incident, we can infer the overarching influence strategies conceptualized in this work.

\subsection{Prevalence of strategies in real influence operations}

\begin{figure*}[t]
    \centering
    \includegraphics[width=\textwidth]{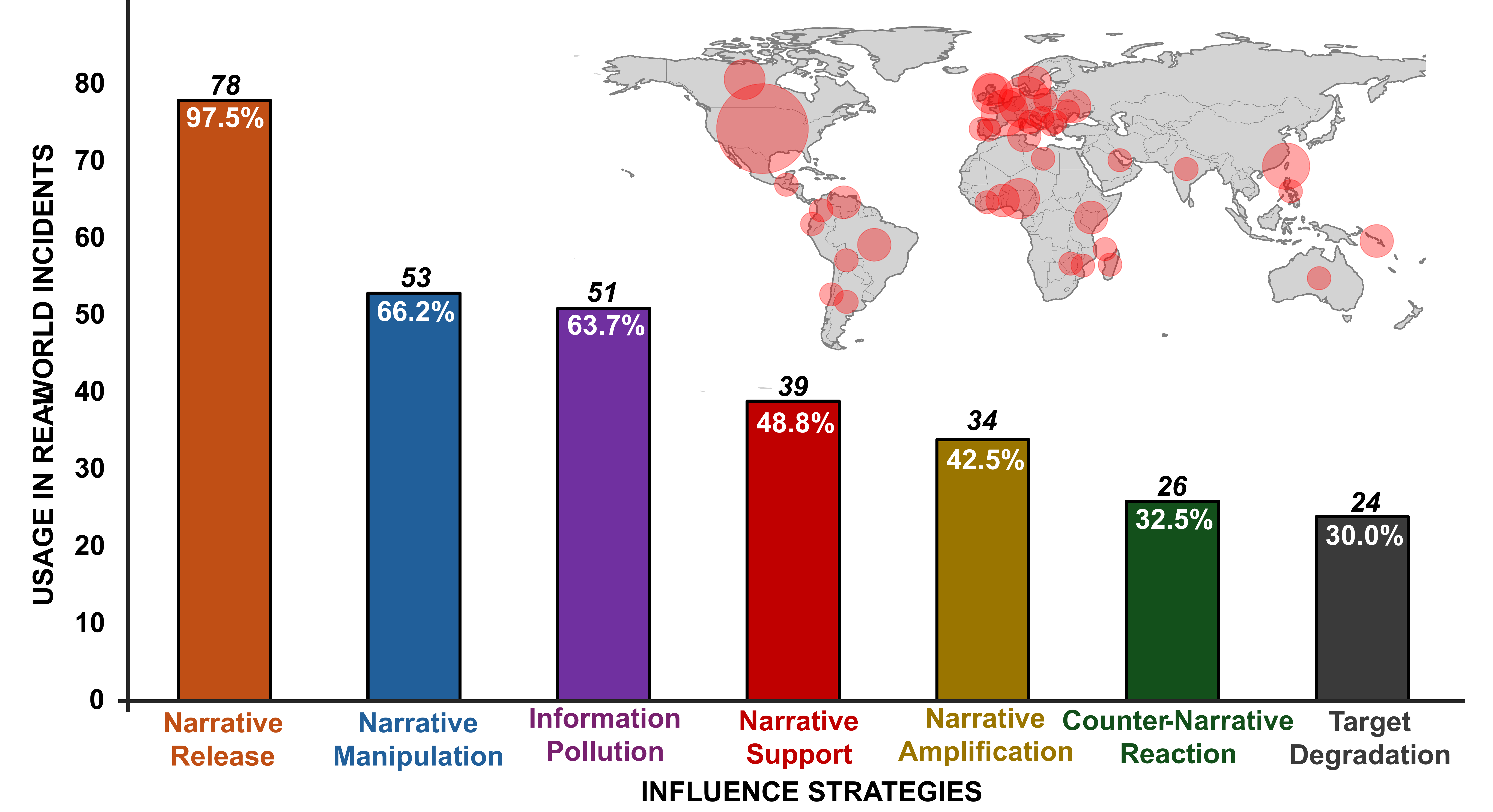}
    \caption{Prevalence of influence strategies in past incidents from 2014 to 2024}
    \label{fig:strategies_distribution}
\end{figure*}

Out of the 81 documented incidents, 80 could be mapped to the influence strategies proposed in this study. Most of them integrating more than one strategy. This indicates that our methodology effectively captures the critical techniques within the complex DISARM framework. Figure~\ref{fig:strategies_distribution} reveals the usage of strategies by adversaries to influence social networks audience.

Among the identified strategies, \textit{Narrative Release} emerges as the most prevalent, present in 97.5\% of cases. This finding underscores the importance of content seeding as the foundation of influence campaigns, where adversaries introduce new narratives, fabricate stories, or selectively leak information to set the stage for broader manipulation efforts. Following closely, \textit{Narrative Manipulation} (66.2\%) and \textit{Information Pollution} (63.7\%) highlight the widespread use of content distortion. These strategies involve reshaping existing narratives, amplifying conspiracy theories, or overwhelming audiences with misleading or conflicting information. Their prevalence suggests that influence operations are not solely about creating falsehoods but also about controlling the information environment to erode trust and manipulate perceptions. \textit{Narrative Support}, observed in 48.8\% of cases, plays a crucial role in reinforcing and sustaining the visibility of manipulated content. Techniques such as coordinated engagement, automated amplification, and deceptive endorsements ensure that seeded narratives gain traction and appear credible. Similarly, \textit{Narrative Amplification} (42.5\%) extends the reach of influence campaigns by leveraging social media algorithms, influencers, and mass-sharing tactics to maximize exposure. Less common but equally significant, \textit{Counter-Narrative Reaction} (32.5\%) and \textit{Target Degradation} (30.0\%) represent more aggressive forms of influence operations.

These findings highlight two key takeaways. First, influence operations extend beyond narrative manipulation (e.g., disinformation), encompassing a broader manipulation of the information ecosystem. Second, real campaigns rarely rely on a single strategy but rather combine them to reinforce and amplify their impact, as inspected next.

\subsection{Dominance of multi-strategy influence operations}
\label{sec:multilayered}

The majority of influence actors (92.5\%, 74 out of 80 cases) rely on multi-strategy operations, rarely deploying a single approach in isolation. The most common tactic (33\%, 24 cases) involves orchestrating four strategies, balancing complexity and efficiency. More sophisticated adversaries execute five-strategy operations (20\%), while three-strategy combinations (19\%) remain a flexible option. Simpler two-strategy operations (15\%) are uncommon, and more intricate campaigns employing six or seven strategies (8\% and 5\%) are rare, likely due to coordination challenges.

\begin{figure*}[t]
    \centering
    \includegraphics[width=0.9\textwidth]{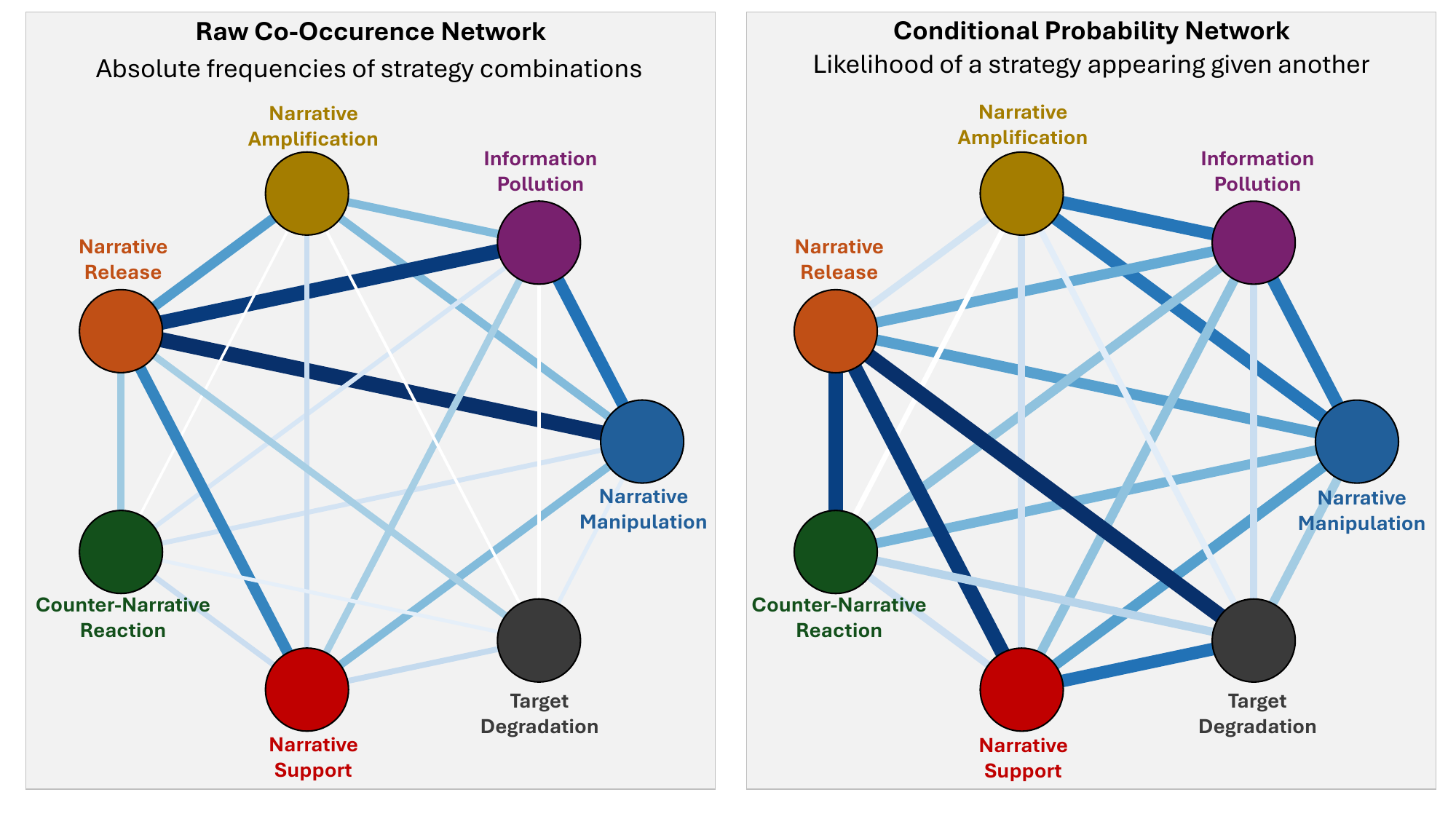}
    \caption{Combination of influence strategies in past incidents from 2014 to 2024}
    \label{fig:strategies_combination}
\end{figure*}

From a total of 120 potential strategy combinations, about 30 distinct patterns emerge from the dataset, revealing preferred multi-strategy playbooks in Figure~\ref{fig:strategies_combination}. On the left, the co-occurrence network illustrates how influence actors interweave different strategies to enhance their impact. A tendency is observed among those who rely on \textit{Narrative Release} as a core strategy, frequently coupling it with \textit{Information Pollution} and \textit{Narrative Manipulation} to both introduce and distort narratives while overwhelming the information space. This structured approach appears in seven incidents. Additionally, some adversaries intensify their manipulation efforts by combining \textit{Information Pollution} and \textit{Narrative Manipulation}, flooding feeds with deceptive content while embedding misleading elements into ongoing discussions. Many further reinforce these efforts by deploying \textit{Narrative Amplification} for wide spread, as observed in 11 cases where all four strategies were used together. Other influence actors rely on \textit{Narrative Support} to strengthen their operations, aligning it with \textit{Narrative Release}, \textit{Narrative Amplification}, \textit{Information Pollution}, and \textit{Narrative Manipulation} to consolidate a coherent influence strategy, appearing together in six operations. While this graph provides insights into dominant playbooks, frequent pairings do not necessarily imply a strong interdependence between strategies.

On the right, the conditional probability network reveals how likely influence actors are to deploy one strategy given the presence of another, uncovering tactical dependencies in influence operations. Two primary trends emerge. First, adversaries who launch \textit{Narrative Release} often combine with \textit{Target Degradation}, \textit{Narrative Support}, and \textit{Counter-Narrative Reaction}, reinforcing their narratives by undermining opposing voices and injecting reactionary content. Second, those relying on \textit{Information Pollution}, \textit{Narrative Manipulation}, and \textit{Narrative Amplification} tend to saturate the information ecosystem with deceptive content, reinforcing misleading narratives through coordinated amplification. Notably, \textit{Narrative Manipulation} appears as a linchpin, frequently used regardless of the strategic combination.

\section{CONFRONTING DIGITAL MANIPULATION: STRATEGIC RESPONSES AND FUTURE DIRECTIONS}

The increasing automation and sophistication of influence operations are transforming them into a persistent digital threat, where adversaries (often APM) deploy coordinated tactics to manipulate narratives, erode trust, and exploit vulnerabilities across social networks~\cite{Fredheim2024}. This study has identified and categorized key influence strategies in social networks, grounded in the DISARM framework. We revealed the prevalence of those strategies through an analysis of 80 FIMI incidents. The co-occurrence and conditional probability perspectives provide complementary insights into how adversaries coordinate multiple strategies, where key narratives are not only introduced and manipulated but also strategically reinforced through amplification and support mechanisms. This analytical and structured approach based on observable strategies in social networks set the stage for a proactive response, integrating cybersecurity methodologies, intelligence-sharing, real-time situational awareness, and mitigation strategies~\cite{Walker2019}.

One of the most promising countermeasures is the integration of Cyber Threat Intelligence (CTI) into influence operation monitoring, allowing analysts to track adversarial tactics and coordinate rapid responses using established intelligence-sharing platforms such as MISP and OpenCTI~\cite{HybridCoE2022}. This structured approach treats disinformation as a cybersecurity threat, fostering real-time collaboration between governments, platforms, and security researchers. However, detection alone is insufficient, and achieving Cyber Situational Awareness (CSA) in social networks is vital to understanding the broader manipulation landscape~\cite{10553682}. By combining OSINT, SIGINT, and HUMINT, and applying AI-driven behavioral analytics, CSA frameworks can anticipate influence campaigns before they escalate, shifting from reactive defense to proactive intervention.

Beyond intelligence and monitoring, risk assessment and mitigation must evolve to counter systemic vulnerabilities~\cite{Walker2019}. Influence operations thrive on media manipulation and institutional weaknesses, making adaptive counter-narratives and targeted disruption strategies essential. The development of AI-driven realistic simulations provides an effective means to evaluate adversary influence, model adversarial behavior, and refine response tactics~\cite{10492674}. These controlled environments enable preemptive stress-testing of emerging disinformation strategies, ensuring that countermeasures are validated before real-world deployment.

Yet, technical solutions alone cannot neutralize influence operations. Education and cognitive resilience remain central to long-term defense. Initiatives like NATO’s InfoRange and Red Team–Blue Team exercises offer practical training in recognizing and countering manipulation tactics~\cite{Walker2019}. Widespread digital literacy programs are needed to prepare both practitioners and the public to critically assess online content~\cite{HybridCoE2022}.

As Generative AI accelerates the automation of information manipulation, the need for cross-disciplinary collaboration becomes more pressing~\cite{blazek2021scotch}. The DISARM framework, serving as a foundation for structuring adversarial influence operations in this work, must continuously adapt to new deception techniques, AI-powered strategies, and cross-platform operations. The future of influence operation mitigation will not rely on a single solution but on a resilient, adaptive ecosystem of intelligence, technology, and education. By embracing this holistic, multi-domain approach, researchers, policymakers, and cybersecurity experts can build more effective defenses against the next generation of digital manipulation threats.

\begin{acks}
This study was partially funded by: (a) the strategic project ``Development of Professionals and Researchers in Cybersecurity, Cyberdefense and Data Science (CDL-TALENTUM)" from i) the Spanish National Institute of Cybersecurity (INCIBE) and ii) by the Recovery, Transformation and Resilience Plan, Next Generation EU; (b) by a ``Juan de la Cierva'' Postdoctoral Fellowship (JDC2023-051658-I) funded by the i) Spanish Ministry of Science, Innovation and Universities (MCIU), ii) by the Spanish State Research Agency (AEI/10.13039/501100011033) and iii) by the European Social Fund Plus (FSE+); (c) by H2020 TMA-MSCA-DN TUAI project "Towards an Understanding of Artificial Intelligence via a transparent, open and explainable perspective" (HORIZON-MSCA-2023-DN-01-01, Grant agreement nº: 101168344); (d) by project PCI2022-134990-2 (MARTINI) of the CHISTERA IV Cofund 2021 program; (e) by EMIF managed by the Calouste Gulbenkian Foundation, in the project MuseAI; and by (f) Comunidad Autonoma de Madrid, and by CIRMA-CAM Project (TEC-2024/COM-404).
\end{acks}

\bibliographystyle{ACM-Reference-Format}
\bibliography{biblio}


\begin{thebibliography}{38}


\ifx \showCODEN    \undefined \def \showCODEN     #1{\unskip}     \fi
\ifx \showDOI      \undefined \def \showDOI       #1{#1}\fi
\ifx \showISBNx    \undefined \def \showISBNx     #1{\unskip}     \fi
\ifx \showISBNxiii \undefined \def \showISBNxiii  #1{\unskip}     \fi
\ifx \showISSN     \undefined \def \showISSN      #1{\unskip}     \fi
\ifx \showLCCN     \undefined \def \showLCCN      #1{\unskip}     \fi
\ifx \shownote     \undefined \def \shownote      #1{#1}          \fi
\ifx \showarticletitle \undefined \def \showarticletitle #1{#1}   \fi
\ifx \showURL      \undefined \def \showURL       {\relax}        \fi
\providecommand\bibfield[2]{#2}
\providecommand\bibinfo[2]{#2}
\providecommand\natexlab[1]{#1}
\providecommand\showeprint[2][]{arXiv:#2}

\bibitem[Alizadeh et~al\mbox{.}(2020)]%
        {Alizadeh2020}
\bibfield{author}{\bibinfo{person}{Meysam Alizadeh}, \bibinfo{person}{Jacob~N. Shapiro}, \bibinfo{person}{Cody Buntain}, {and} \bibinfo{person}{Joshua~A. Tucker}.} \bibinfo{year}{2020}\natexlab{}.
\newblock \showarticletitle{{Content-based features predict social media influence operations}}.
\newblock \bibinfo{journal}{\emph{Science Advances}} \bibinfo{volume}{6}, \bibinfo{number}{30} (\bibinfo{year}{2020}), \bibinfo{pages}{1--13}.
\newblock
\showISSN{23752548}
\urldef\tempurl%
\url{https://doi.org/10.1126/sciadv.abb5824}
\showDOI{\tempurl}


\bibitem[{ATHENEA Project}(2024)]%
        {athenea}
\bibfield{author}{\bibinfo{person}{{ATHENEA Project}}.} \bibinfo{year}{2024}\natexlab{}.
\newblock \bibinfo{title}{Policy Brief Conclusions and Recommendations from the ATHENA Project on Foreign Information Manipulation and Interference}.
\newblock
\newblock
\urldef\tempurl%
\url{https://project-athena.eu/wp-content/uploads/2024/12/Policy-Brief-ATHENA.pdf}
\showURL{%
\tempurl}


\bibitem[Bergh(2020)]%
        {bergh2020understanding}
\bibfield{author}{\bibinfo{person}{Arild Bergh}.} \bibinfo{year}{2020}\natexlab{}.
\newblock \showarticletitle{Understanding influence operations in social media}.
\newblock \bibinfo{journal}{\emph{Journal of information warfare}} \bibinfo{volume}{19}, \bibinfo{number}{4} (\bibinfo{year}{2020}), \bibinfo{pages}{110--131}.
\newblock
\urldef\tempurl%
\url{https://www.jstor.org/stable/27033648}
\showURL{%
\tempurl}


\bibitem[Blazek(2021)]%
        {blazek2021scotch}
\bibfield{author}{\bibinfo{person}{Sam Blazek}.} \bibinfo{year}{2021}\natexlab{}.
\newblock \showarticletitle{SCOTCH: A framework for rapidly assessing influence operations}.
\newblock \bibinfo{journal}{\emph{Atlantic Council}} (\bibinfo{year}{2021}).
\newblock
\urldef\tempurl%
\url{https://www.atlanticcouncil.org/blogs/geotech-cues/scotch-a-framework-for-rapidly-assessing-influence-operations}
\showURL{%
\tempurl}


\bibitem[Blumenstock et~al\mbox{.}(2025)]%
        {blumenstock2025migration}
\bibfield{author}{\bibinfo{person}{Joshua~E Blumenstock}, \bibinfo{person}{Guanghua Chi}, {and} \bibinfo{person}{Xu Tan}.} \bibinfo{year}{2025}\natexlab{}.
\newblock \showarticletitle{Migration and the value of social networks}.
\newblock \bibinfo{journal}{\emph{Review of Economic Studies}} \bibinfo{volume}{92}, \bibinfo{number}{1} (\bibinfo{year}{2025}), \bibinfo{pages}{97--128}.
\newblock
\urldef\tempurl%
\url{https://doi.org/10.1093/restud/rdad113}
\showDOI{\tempurl}


\bibitem[Buchanan et~al\mbox{.}(2021)]%
        {buchanan2021truth}
\bibfield{author}{\bibinfo{person}{Ben Buchanan}, \bibinfo{person}{Andrew Lohn}, \bibinfo{person}{Micah Musser}, {and} \bibinfo{person}{Katerina Sedova}.} \bibinfo{year}{2021}\natexlab{}.
\newblock \showarticletitle{Truth, lies, and automation}.
\newblock \bibinfo{journal}{\emph{Center for Security and Emerging technology}} \bibinfo{volume}{1}, \bibinfo{number}{1} (\bibinfo{year}{2021}), \bibinfo{pages}{2}.
\newblock


\bibitem[Budak et~al\mbox{.}(2024)]%
        {Budak2024}
\bibfield{author}{\bibinfo{person}{Ceren Budak}, \bibinfo{person}{Brendan Nyhan}, \bibinfo{person}{David~M. Rothschild}, \bibinfo{person}{Emily Thorson}, {and} \bibinfo{person}{Duncan~J. Watts}.} \bibinfo{year}{2024}\natexlab{}.
\newblock \showarticletitle{{Misunderstanding the harms of online misinformation}}.
\newblock \bibinfo{journal}{\emph{Nature}} \bibinfo{volume}{630}, \bibinfo{number}{8015} (\bibinfo{year}{2024}), \bibinfo{pages}{45--53}.
\newblock
\showISSN{14764687}
\urldef\tempurl%
\url{https://doi.org/10.1038/s41586-024-07417-w}
\showDOI{\tempurl}


\bibitem[C{\'a}novas L{\'o}pez~de Molina et~al\mbox{.}(2024)]%
        {canovas2024analyzing}
\bibfield{author}{\bibinfo{person}{Gonzalo C{\'a}novas L{\'o}pez~de Molina}, \bibinfo{person}{Felipe S{\'a}nchez~Gonz{\'a}lez}, \bibinfo{person}{Pantaleone Nespoli}, \bibinfo{person}{Javier Pastor~Galindo}, {and} \bibinfo{person}{Jos{\'e}~A Ruip{\'e}rez~Valiente}.} \bibinfo{year}{2024}\natexlab{}.
\newblock \showarticletitle{Analyzing frameworks to model disinformation attacks in online social networks}. In \bibinfo{booktitle}{\emph{Proceedings of the 9th National Conference on Cybersecurity Research (JNIC2024)}}. \bibinfo{pages}{92--99}.
\newblock


\bibitem[Carley(2020)]%
        {carley2020social}
\bibfield{author}{\bibinfo{person}{Kathleen~M Carley}.} \bibinfo{year}{2020}\natexlab{}.
\newblock \showarticletitle{Social cybersecurity: an emerging science}.
\newblock \bibinfo{journal}{\emph{Computational and mathematical organization theory}} \bibinfo{volume}{26}, \bibinfo{number}{4} (\bibinfo{year}{2020}), \bibinfo{pages}{365--381}.
\newblock
\urldef\tempurl%
\url{https://doi.org/10.1007/s10588-020-09322-9}
\showDOI{\tempurl}


\bibitem[Cavaliere et~al\mbox{.}(2024)]%
        {10553682}
\bibfield{author}{\bibinfo{person}{Danilo Cavaliere}, \bibinfo{person}{Giuseppe Fenza}, \bibinfo{person}{Domenico Furno}, {and} \bibinfo{person}{Vincenzo Loia}.} \bibinfo{year}{2024}\natexlab{}.
\newblock \showarticletitle{A semantic model bridging DISARM framework and Situation Awareness for disinformation Attacks Attribution}. In \bibinfo{booktitle}{\emph{2024 IEEE Conference on Cognitive and Computational Aspects of Situation Management (CogSIMA)}}. \bibinfo{pages}{55--62}.
\newblock
\urldef\tempurl%
\url{https://doi.org/10.1109/CogSIMA61085.2024.10553682}
\showDOI{\tempurl}


\bibitem[Desouza et~al\mbox{.}(2020)]%
        {DESOUZA2020101606}
\bibfield{author}{\bibinfo{person}{Kevin~C. Desouza}, \bibinfo{person}{Atif Ahmad}, \bibinfo{person}{Humza Naseer}, {and} \bibinfo{person}{Munish Sharma}.} \bibinfo{year}{2020}\natexlab{}.
\newblock \showarticletitle{Weaponizing information systems for political disruption: The Actor, Lever, Effects, and Response Taxonomy (ALERT)}.
\newblock \bibinfo{journal}{\emph{Computers \& Security}}  \bibinfo{volume}{88} (\bibinfo{year}{2020}), \bibinfo{pages}{101606}.
\newblock
\showISSN{0167-4048}
\urldef\tempurl%
\url{https://doi.org/10.1016/j.cose.2019.101606}
\showDOI{\tempurl}


\bibitem[{European Union External Action}(2024)]%
        {EuropeanUnionExternalAction2024}
\bibfield{author}{\bibinfo{person}{{European Union External Action}}.} \bibinfo{year}{2024}\natexlab{}.
\newblock \bibinfo{booktitle}{\emph{{2nd EEAS Report on Foreign Information Manipulation and Interference Threats}}}.
\newblock \bibinfo{type}{{T}echnical {R}eport} January.
\newblock
\urldef\tempurl%
\url{https://www.eeas.europa.eu/eeas/2nd-eeas-report-foreign-information-manipulation-and-interference-threats_en}
\showURL{%
\tempurl}


\bibitem[for Network and Security(2014)]%
        {enisathreat}
\bibfield{author}{\bibinfo{person}{European Union~Agency for Network} {and} \bibinfo{person}{Information Security}.} \bibinfo{year}{2014}\natexlab{}.
\newblock \bibinfo{title}{ENISA Threat Landscape 2024}.
\newblock
\newblock
Issue December.
\showISBNx{9789292046750}
\urldef\tempurl%
\url{https://www.enisa.europa.eu/publications/enisa-threat-landscape-2024}
\showURL{%
\tempurl}


\bibitem[Fredheim and Pamment(2024)]%
        {Fredheim2024}
\bibfield{author}{\bibinfo{person}{Rolf Fredheim} {and} \bibinfo{person}{James Pamment}.} \bibinfo{year}{2024}\natexlab{}.
\newblock \showarticletitle{Assessing the risks and opportunities posed by AI-enhanced influence operations on social media}.
\newblock \bibinfo{journal}{\emph{Place Branding and Public Diplomacy}} (\bibinfo{year}{2024}), \bibinfo{pages}{1--8}.
\newblock
\urldef\tempurl%
\url{https://doi.org/10.1057/s41254-023-00322-5}
\showDOI{\tempurl}


\bibitem[Fulde-Hardy(2024)]%
        {FuldeHardy2024}
\bibfield{author}{\bibinfo{person}{Margot Fulde-Hardy}.} \bibinfo{year}{2024}\natexlab{}.
\newblock \bibinfo{booktitle}{\emph{Foreign Information Manipulation and Interference (FIMI) Activities Targeting Elections (2014-2024): Presenting a dataset, a methodology, and a codebook to guide future applications of structured frameworks enabling threat assessment}}.
\newblock \bibinfo{type}{{T}echnical {R}eport}. \bibinfo{institution}{Working Paper}.
\newblock
\urldef\tempurl%
\url{https://igp.sipa.columbia.edu/sites/igp/files/2024-06/2024_06_07_FIMIElections_Margot_Fulde-Hardy_paper.pdf}
\showURL{%
\tempurl}


\bibitem[Gabriel et~al\mbox{.}(2023)]%
        {gabriel2023inductive}
\bibfield{author}{\bibinfo{person}{Nicholas~A Gabriel}, \bibinfo{person}{David~A Broniatowski}, {and} \bibinfo{person}{Neil~F Johnson}.} \bibinfo{year}{2023}\natexlab{}.
\newblock \showarticletitle{Inductive detection of influence operations via graph learning}.
\newblock \bibinfo{journal}{\emph{Scientific Reports}} \bibinfo{volume}{13}, \bibinfo{number}{1} (\bibinfo{year}{2023}), \bibinfo{pages}{22571}.
\newblock
\urldef\tempurl%
\url{https://doi.org/10.1038/s41598-023-49676-z}
\showDOI{\tempurl}


\bibitem[Goldstein et~al\mbox{.}(2023)]%
        {Goldstein2023}
\bibfield{author}{\bibinfo{person}{Josh~A. Goldstein}, \bibinfo{person}{Girish Sastry}, \bibinfo{person}{Micah Musser}, \bibinfo{person}{Renee DiResta}, \bibinfo{person}{Matthew Gentzel}, {and} \bibinfo{person}{Katerina Sedova}.} \bibinfo{year}{2023}\natexlab{}.
\newblock \bibinfo{booktitle}{\emph{{Generative Language Models and Automated Influence Operations: Emerging Threats and Potential Mitigations}}}.
\newblock \bibinfo{type}{{T}echnical {R}eport}. \bibinfo{institution}{OpenAI}.
\newblock


\bibitem[Group(2025)]%
        {Google2025}
\bibfield{author}{\bibinfo{person}{Google Threat~Intelligence Group}.} \bibinfo{year}{2025}\natexlab{}.
\newblock \bibinfo{booktitle}{\emph{Adversarial Misuse of Generative AI}}.
\newblock \bibinfo{type}{{T}echnical {R}eport}. \bibinfo{institution}{Google}.
\newblock


\bibitem[H{\'e}nin(2023)]%
        {disinfolabeufimi}
\bibfield{author}{\bibinfo{person}{Nicolas H{\'e}nin}.} \bibinfo{year}{2023}\natexlab{}.
\newblock \showarticletitle{FIMI: Towards a European Redefinition of Foreign Interference}.
\newblock \bibinfo{journal}{\emph{EU DISINFO Lab}}  \bibinfo{volume}{13} (\bibinfo{year}{2023}), \bibinfo{pages}{2023}.
\newblock


\bibitem[Hristakieva et~al\mbox{.}(2022)]%
        {hristakieva2022spread}
\bibfield{author}{\bibinfo{person}{Kristina Hristakieva}, \bibinfo{person}{Stefano Cresci}, \bibinfo{person}{Giovanni Da~San~Martino}, \bibinfo{person}{Mauro Conti}, {and} \bibinfo{person}{Preslav Nakov}.} \bibinfo{year}{2022}\natexlab{}.
\newblock \showarticletitle{The spread of propaganda by coordinated communities on social media}. In \bibinfo{booktitle}{\emph{Proceedings of the 14th ACM Web Science Conference 2022}}. \bibinfo{pages}{191--201}.
\newblock
\urldef\tempurl%
\url{https://doi.org/10.1145/3501247.3531543}
\showDOI{\tempurl}


\bibitem[Kaiser et~al\mbox{.}(2022)]%
        {kaiser2022partisan}
\bibfield{author}{\bibinfo{person}{Johannes Kaiser}, \bibinfo{person}{Cristian Vaccari}, {and} \bibinfo{person}{Andrew Chadwick}.} \bibinfo{year}{2022}\natexlab{}.
\newblock \showarticletitle{Partisan blocking: Biased responses to shared misinformation contribute to network polarization on social media}.
\newblock \bibinfo{journal}{\emph{Journal of Communication}} \bibinfo{volume}{72}, \bibinfo{number}{2} (\bibinfo{year}{2022}), \bibinfo{pages}{214--240}.
\newblock
\urldef\tempurl%
\url{https://doi.org/10.1093/joc/jqac002}
\showDOI{\tempurl}


\bibitem[Microsoft(2024)]%
        {Microsoft2024}
\bibfield{author}{\bibinfo{person}{Microsoft}.} \bibinfo{year}{2024}\natexlab{}.
\newblock \bibinfo{booktitle}{\emph{Microsoft Digital Defense Report 2024: The Foundations and New Frontiers of Cybersecurity}}.
\newblock \bibinfo{type}{{T}echnical {R}eport}. \bibinfo{institution}{Microsoft}.
\newblock


\bibitem[Mirza et~al\mbox{.}(2023)]%
        {mirza2023tactics}
\bibfield{author}{\bibinfo{person}{Muhammad~Shujaat Mirza}, \bibinfo{person}{Labeeba Begum}, \bibinfo{person}{Liang Niu}, \bibinfo{person}{Sarah Pardo}, \bibinfo{person}{Azza Abouzied}, \bibinfo{person}{Paolo Papotti}, {and} \bibinfo{person}{Christina P{\"o}pper}.} \bibinfo{year}{2023}\natexlab{}.
\newblock \showarticletitle{Tactics, Threats \& Targets: Modeling Disinformation and its Mitigation}. In \bibinfo{booktitle}{\emph{NDSS}}.
\newblock
\urldef\tempurl%
\url{https://doi.org/10.14722/ndss.2023.23657}
\showDOI{\tempurl}


\bibitem[Nimmo and Hutchins(2023)]%
        {nimmo2023phase}
\bibfield{author}{\bibinfo{person}{Ben Nimmo} {and} \bibinfo{person}{Eric Hutchins}.} \bibinfo{year}{2023}\natexlab{}.
\newblock \bibinfo{booktitle}{\emph{Phase-based tactical analysis of online operations}}.
\newblock \bibinfo{type}{{T}echnical {R}eport}. \bibinfo{institution}{Carnegie Endowment for International Peace}.
\newblock
\urldef\tempurl%
\url{https://www.jstor.org/stable/resrep48480.7}
\showURL{%
\tempurl}


\bibitem[Pamment(2020)]%
        {pamment2020eu}
\bibfield{author}{\bibinfo{person}{James Pamment}.} \bibinfo{year}{2020}\natexlab{}.
\newblock \bibinfo{booktitle}{\emph{The EU's Role in Fighting Disinformation: Crafting A Disinformation Framework}}.
\newblock \bibinfo{publisher}{Carnegie Endowment for International Peace.}
\newblock
\urldef\tempurl%
\url{https://www.jstor.org/stable/resrep26180.1}
\showURL{%
\tempurl}


\bibitem[Pan et~al\mbox{.}(2023)]%
        {pan-etal-2023-risk}
\bibfield{author}{\bibinfo{person}{Yikang Pan}, \bibinfo{person}{Liangming Pan}, \bibinfo{person}{Wenhu Chen}, \bibinfo{person}{Preslav Nakov}, \bibinfo{person}{Min-Yen Kan}, {and} \bibinfo{person}{William Wang}.} \bibinfo{year}{2023}\natexlab{}.
\newblock \showarticletitle{On the Risk of Misinformation Pollution with Large Language Models}. In \bibinfo{booktitle}{\emph{Findings of the Association for Computational Linguistics: EMNLP 2023}}, \bibfield{editor}{\bibinfo{person}{Houda Bouamor}, \bibinfo{person}{Juan Pino}, {and} \bibinfo{person}{Kalika Bali}} (Eds.). \bibinfo{publisher}{Association for Computational Linguistics}, \bibinfo{address}{Singapore}, \bibinfo{pages}{1389--1403}.
\newblock
\urldef\tempurl%
\url{https://doi.org/10.18653/v1/2023.findings-emnlp.97}
\showDOI{\tempurl}


\bibitem[Panizio(2023)]%
        {edmoelections}
\bibfield{author}{\bibinfo{person}{Enzo Panizio}.} \bibinfo{year}{2023}\natexlab{}.
\newblock \bibinfo{booktitle}{\emph{Disinformation narratives during the 2023 elections in Europe}}.
\newblock \bibinfo{type}{{T}echnical {R}eport}. \bibinfo{institution}{European Digital Media Observatory (EDMO)}.
\newblock
Issue November 2023.
\urldef\tempurl%
\url{https://edmo.eu/publications/disinformation-narratives-during-the-2023-elections-in-europe}
\showURL{%
\tempurl}


\bibitem[Pastor-Galindo et~al\mbox{.}(2022)]%
        {PASTORGALINDO2022161}
\bibfield{author}{\bibinfo{person}{Javier Pastor-Galindo}, \bibinfo{person}{Félix {Gómez Mármol}}, {and} \bibinfo{person}{Gregorio {Martínez Pérez}}.} \bibinfo{year}{2022}\natexlab{}.
\newblock \showarticletitle{Profiling users and bots in Twitter through social media analysis}.
\newblock \bibinfo{journal}{\emph{Information Sciences}}  \bibinfo{volume}{613} (\bibinfo{year}{2022}), \bibinfo{pages}{161--183}.
\newblock
\showISSN{0020-0255}
\urldef\tempurl%
\url{https://doi.org/10.1016/j.ins.2022.09.046}
\showDOI{\tempurl}


\bibitem[Pastor-Galindo et~al\mbox{.}(2021)]%
        {9451574}
\bibfield{author}{\bibinfo{person}{Javier Pastor-Galindo}, \bibinfo{person}{Félix~Gómez Mármol}, {and} \bibinfo{person}{Gregorio~Martínez Pérez}.} \bibinfo{year}{2021}\natexlab{}.
\newblock \showarticletitle{Nothing to Hide? On the Security and Privacy Threats Beyond Open Data}.
\newblock \bibinfo{journal}{\emph{IEEE Internet Computing}} \bibinfo{volume}{25}, \bibinfo{number}{4} (\bibinfo{year}{2021}), \bibinfo{pages}{58--66}.
\newblock
\urldef\tempurl%
\url{https://doi.org/10.1109/MIC.2021.3088335}
\showDOI{\tempurl}


\bibitem[Pastor-Galindo et~al\mbox{.}(2024)]%
        {10492674}
\bibfield{author}{\bibinfo{person}{Javier Pastor-Galindo}, \bibinfo{person}{Pantaleone Nespoli}, {and} \bibinfo{person}{José~A. Ruipérez-Valiente}.} \bibinfo{year}{2024}\natexlab{}.
\newblock \showarticletitle{Large-Language-Model-Powered Agent-Based Framework for Misinformation and Disinformation Research: Opportunities and Open Challenges}.
\newblock \bibinfo{journal}{\emph{IEEE Security \& Privacy}} \bibinfo{volume}{22}, \bibinfo{number}{3} (\bibinfo{year}{2024}), \bibinfo{pages}{24--36}.
\newblock
\urldef\tempurl%
\url{https://doi.org/10.1109/MSEC.2024.3380511}
\showDOI{\tempurl}


\bibitem[{R. Walker} et~al\mbox{.}(2019)]%
        {Walker2019}
\bibfield{author}{\bibinfo{person}{Christopher {R. Walker}}, \bibinfo{person}{Sara-Jayne Terp}, \bibinfo{person}{Pablo {C. Breuer}}, {and} \bibinfo{person}{Courtney {L. Crooks, PhD}}.} \bibinfo{year}{2019}\natexlab{}.
\newblock \showarticletitle{{Misinfosec: Applying Information Security Paradigms to Misinformation Campaigns}}.
\newblock  (\bibinfo{year}{2019}), \bibinfo{pages}{1026--1032}.
\newblock
\showISBNx{9781450366755}
\urldef\tempurl%
\url{https://doi.org/10.1145/3308560.3316742}
\showDOI{\tempurl}


\bibitem[Ren et~al\mbox{.}(2022)]%
        {ren2022authoritarian}
\bibfield{author}{\bibinfo{person}{Zhiying~Bella Ren}, \bibinfo{person}{Andrew~M Carton}, \bibinfo{person}{Eugen Dimant}, {and} \bibinfo{person}{Maurice~E Schweitzer}.} \bibinfo{year}{2022}\natexlab{}.
\newblock \showarticletitle{Authoritarian leaders share conspiracy theories to attack opponents, galvanize followers, shift blame, and undermine democratic institutions}.
\newblock \bibinfo{journal}{\emph{Current Opinion in Psychology}}  \bibinfo{volume}{46} (\bibinfo{year}{2022}), \bibinfo{pages}{101388}.
\newblock
\urldef\tempurl%
\url{https://doi.org/10.1016/j.copsyc.2022.101388}
\showDOI{\tempurl}


\bibitem[Sikora and Schafer(2025)]%
        {reportasd}
\bibfield{author}{\bibinfo{person}{Krystyna Sikora} {and} \bibinfo{person}{Bret Schafer}.} \bibinfo{year}{2025}\natexlab{}.
\newblock \bibinfo{booktitle}{\emph{{The State(s) of Foreign Information Operations. A State-by-State Look at Foreign Information Manipulation in the United States}}}.
\newblock \bibinfo{type}{{T}echnical {R}eport}. \bibinfo{institution}{Alliance for Securing Democracy}.
\newblock
\urldef\tempurl%
\url{https://securingdemocracy.gmfus.org/wp-content/uploads/2025/02/The-States-of-Foreign-Information-Operations.pdf}
\showURL{%
\tempurl}


\bibitem[Sixto and Kim(2023)]%
        {Sixto2023}
\bibfield{author}{\bibinfo{person}{Daniel~R Sixto} {and} \bibinfo{person}{Paul~S Kim}.} \bibinfo{year}{2023}\natexlab{}.
\newblock \bibinfo{booktitle}{\emph{{Structured Process for Information Campaign Enhancement (SP!CE) 2.1 An Analytic Framework, Knowledge Base, and Scoring Rubric for Operations in the Information Environment McLean, VA}}}.
\newblock \bibinfo{type}{{T}echnical {R}eport} 0704.
\newblock
\urldef\tempurl%
\url{https://apps.dtic.mil/sti/trecms/pdf/AD1214088.pdf}
\showURL{%
\tempurl}


\bibitem[Smith et~al\mbox{.}(2024)]%
        {adacdisarm}
\bibfield{author}{\bibinfo{person}{Victoria Smith}, \bibinfo{person}{Stephen Campbell}, {and} \bibinfo{person}{Adam Maunder}.} \bibinfo{year}{2024}\natexlab{}.
\newblock \bibinfo{title}{A Comprehensive Review of DISARM Framework and its Compatibility with Related Frameworks Used to Model Foreign Information Manipulation and Interference}.
\newblock
\newblock
\urldef\tempurl%
\url{https://adacio.eu/disarm-august2024}
\showURL{%
\tempurl}


\bibitem[Terp and Breuer(2022)]%
        {terp2022disarm}
\bibfield{author}{\bibinfo{person}{Sara-Jayne Terp} {and} \bibinfo{person}{Pablo Breuer}.} \bibinfo{year}{2022}\natexlab{}.
\newblock \showarticletitle{Disarm: a framework for analysis of disinformation campaigns}. In \bibinfo{booktitle}{\emph{2022 IEEE Conference on Cognitive and Computational Aspects of Situation Management (CogSIMA)}}. IEEE, \bibinfo{pages}{1--8}.
\newblock
\urldef\tempurl%
\url{https://doi.org/10.1109/CogSIMA54611.2022.9830669}
\showDOI{\tempurl}


\bibitem[{The European Centre of Excellence for Countering Hybrid Threats (Hybrid CoE)}(2022)]%
        {HybridCoE2022}
\bibfield{author}{\bibinfo{person}{{The European Centre of Excellence for Countering Hybrid Threats (Hybrid CoE)}}.} \bibinfo{year}{2022}\natexlab{}.
\newblock \bibinfo{booktitle}{\emph{{Foreign information manipulation and interference defence standards: Test for rapid adoption of the common language and framework 'DISARM' The European Centre of Excellence for Countering Hybrid Threats}}}.
\newblock
\showISBNx{9789527472460}
\urldef\tempurl%
\url{https://www.hybridcoe.fi/wp-content/uploads/2022/11/20221129_Hybrid_CoE_Research_Report_7_Disarm_WEB.pdf}
\showURL{%
\tempurl}


\bibitem[Tyushka(2022)]%
        {tyushka2022weaponizing}
\bibfield{author}{\bibinfo{person}{Andriy Tyushka}.} \bibinfo{year}{2022}\natexlab{}.
\newblock \showarticletitle{Weaponizing narrative: Russia contesting EUrope’s liberal identity, power and hegemony}.
\newblock \bibinfo{journal}{\emph{Journal of Contemporary European Studies}} \bibinfo{volume}{30}, \bibinfo{number}{1} (\bibinfo{year}{2022}), \bibinfo{pages}{115--135}.
\newblock
\urldef\tempurl%
\url{https://doi.org/10.1080/14782804.2021.1883561}
\showDOI{\tempurl}


\end{thebibliography}

\end{document}